\documentclass{DISproc}

\begin{document}
\title{Fully massive Scheme for Jet Production in DIS }

\author{{\slshape Piotr Kotko, Wojciech Slominski}\\[1ex]
M. Smoluchowski Institute of Physics, Jagiellonian Univ., Cracow\footnote{The work supported by the Polish National Science Centre grant No. 
DEC-2011/01/B/ST2/03643.} }

\contribID{xy}

\doi  

\maketitle

\begin{abstract}
We present a consistent treatment of heavy quarks for jet production
 in DIS at NLO accuracy. The method is based on the ACOT
massive factorization scheme and dipole subtraction method for jets.
The last had to be however extended in order to take into account
initial state splittings with heavy quarks. We constructed relevant
kinematics and dipole splitting functions together with their integrals.
We partially implemented the method in a MC program and checked against
the known inclusive result for charm structure function.
\end{abstract}

\section{Introduction}

There are two basic approaches to heavy quarks production in DIS.
First is so called zero-mass variable flavor number scheme (ZM-VFNS), which
treats heavy quarks as massless partons with corresponding parton
distribution functions (PDF). This scheme is applicable when the hard
scale (taken here as the virtuality of the exchanged boson $Q^{2}$)
is much larger then the mass $m_{\mathbf{Q}}$ of a given heavy quark $\mathbf{Q}$.
On the other hand, when $Q^{2}$ is of the order of $m_{\mathbf{Q}}$,
so called fixed flavor number scheme (FFNS) is applicable. It retains
the full mass dependence in the coefficient function and there is no
PDF for $\mathbf{Q}$ as to the leading power it cannot appear
in the soft part.

Increasing precision of the data forces us to control also the intermediate
region of $Q^{2}$. The methods that address this problem are called
general mass schemes (GM) \cite{Aivazis:1993pi,Buza:1996wv,Thorne:1997ga,Forte:2010ta}.
They are however formulated for inclusive processes only and similar
method relevant for jets is highly desirable.

In the following we briefly describe our solution to this problem
\cite{Kotko_phdthesis,Kotko:2012ws}. It is based on the ACOT massive factorization theorem \cite{Aivazis:1993pi,Collins:1998rz}
and massive dipole subtraction method (DSM) \cite{Dittmaier:1999mb,Phaf:2001gc,Catani:2002hc},
which however had to be reformulated in order to match with the former.

\section{Dipole subtraction method with massive partons}

Consider NLO calculation of a cross section for producing $n$ jets
in lepton-hadron reaction. The LO cross section is schematically written
as\begin{equation}
\sigma_{n}^{\left(\mathrm{LO}\right)}=\mathcal{N}\,\sum_{a}f_{a}\otimes\,\int d\Phi_{n,a}\,\left|\mathcal{M}_{n,a}\right|^{2}F_{n,a},\end{equation}
where $\mathcal{N}$ is a normalization factor, $f_{a}$ are PDFs, $d\Phi_{n,a}$
is $n$-parton phase space (PS) and $\mathcal{M}_{n,a}$ is a tree-level
matrix element (ME) with $n$ final state partons and one QCD initial state parton
$a$. The jet function $F_{n,a}$ determines
the actual observable and is realized by a suitable jet algorithm. It
satisfies $F_{n+1,a}=F_{n,a}$ in the singular regions of PS. At NLO
the corrections involve loop diagrams living on $n$-particle PS and
additional real emission belonging to $\left(n+1\right)$-particle
PS. Both contain IR singularities which ultimately cancel, however
the cancellation is non-trivial as the singularities have different
origin. An elegant and exact solution is provided by DSM. One adds
and subtracts an auxiliary contribution $\mathcal{D}_{n,a}$, such that it mimics all the singularities of $\mathcal{M}_{n+1,a}$
and at the same time can be analytically integrated over singular
regions of PS. To be more specific we have\begin{multline}
\sigma_{n}^{\left(\mathrm{NLO}\right)}=\mathcal{N}\,\sum_{a}f_{a}\otimes\Bigg\{\int d\Phi_{n+1,a}\,\left[\left|\mathcal{M}_{n+1,a}\right|^{2}F_{n+1,a}-\mathcal{D}_{n,a}F_{n,a}\right]\\
+\int d\Phi_{n,a}\,\left[\mathcal{M}_{n,a}^{\left(\mathrm{loop}\right)}+\int d\phi_{a}\,\mathcal{D}_{n,a}-\mathcal{C}_{n,a}\right]\, F_{n,a}\Bigg\},\end{multline}
where virtual corrections to $\left|\mathcal{M}_{n,a}\right|^{2}$
are symbolically denoted as $\mathcal{M}_{n,a}^{\left(\mathrm{loop}\right)}$.
The subspace leading to singularities is $d\phi_{a}$ and
fulfils PS factorization formula $d\Phi_{n+1,a}=d\Phi_{n,a}\otimes d\phi_{a}$.
Thanks to the properties of $\mathcal{D}_{n,a}$ and $F_{n+1,a}$ the first square bracket is integrable in four
dimensions, while in the second, the poles resulting from integral
$\int d\phi_{a}\,\mathcal{D}_{n,a}$ are cancelled against the ones
in $\mathcal{M}_{n,a}^{\left(\mathrm{loop}\right)}$. However, not
all singularities cancel in this way -- there are also collinear poles
connected with initial state splittings of massless partons. They
are removed by means of collinear subtraction term $\mathcal{C}_{n,a}$.
It has the form\begin{equation}
\mathcal{C}_{n,a}=\sum_{b}\mathcal{F}_{ab}\otimes\left|\mathcal{M}_{n,b}\right|^{2},\label{eq:coll_sub_term}\end{equation}
where $\mathcal{F}_{ab}$ are renormalized partonic PDFs, i.e. the
densities of partons $b$ inside parton $a$. For instance in the $\overline{\mathrm{MS}}$
scheme $\mathcal{F}_{ab}\left(z\right)=-\frac{1}{\varepsilon}\,\frac{\alpha_{s}}{2\pi}\, P_{ab}\left(z\right)$,
where $P_{ab}\left(z\right)$ are standard splitting functions.

Within DSM the dipole function is realized as a sum of contributions
corresponding to single emissions with different combinations of ``emitter'' and ``spectator''%
partons\footnote{The notion ``emitter'' and ``spectator'' are explained in \cite{Catani:1996vz}}.
Each such term $D$ has a general form\begin{equation}
D=\hat{V}\,\hat{C}\left|\hat{\mathcal{M}}_{n,a}\right|^{2},\label{eq:dipole}\end{equation}
where $\hat{V}$ is so called dipole splitting matrix (in helicity
space) and encodes the information about some of the singularities
of $\mathcal{M}_{n+1,a}$. The matrix $\hat{C}$ corresponds to
color operators for parton emissions, which act on the matrix element. The notation above is
symbolic and means that both quantities are correlated in the color and
spin space.
For DIS, there are three
different classes of dipoles $D$, depending on whether emitter or
spectator are in the initial or final states. Here we are mainly interested
in the case of initial state emitter and final state spectator as
they contain factorization-related information.

When heavy quarks are present, the above general picture remains the same.
If however a massive parton takes part in a splitting process there
is no collinear singularity. Nevertheless there are IR sensitive logarithms
which become harmful when the external scale becomes large. We
shall refer to such terms as \textit{quasi-collinear singularities} \cite{Catani:2000ef}
and abbreviate as q-singularities.

\begin{wrapfigure}[17]{r}{0.5\textwidth}%
\vspace{-10pt}
\begin{centering}
\includegraphics[width=0.5\textwidth]{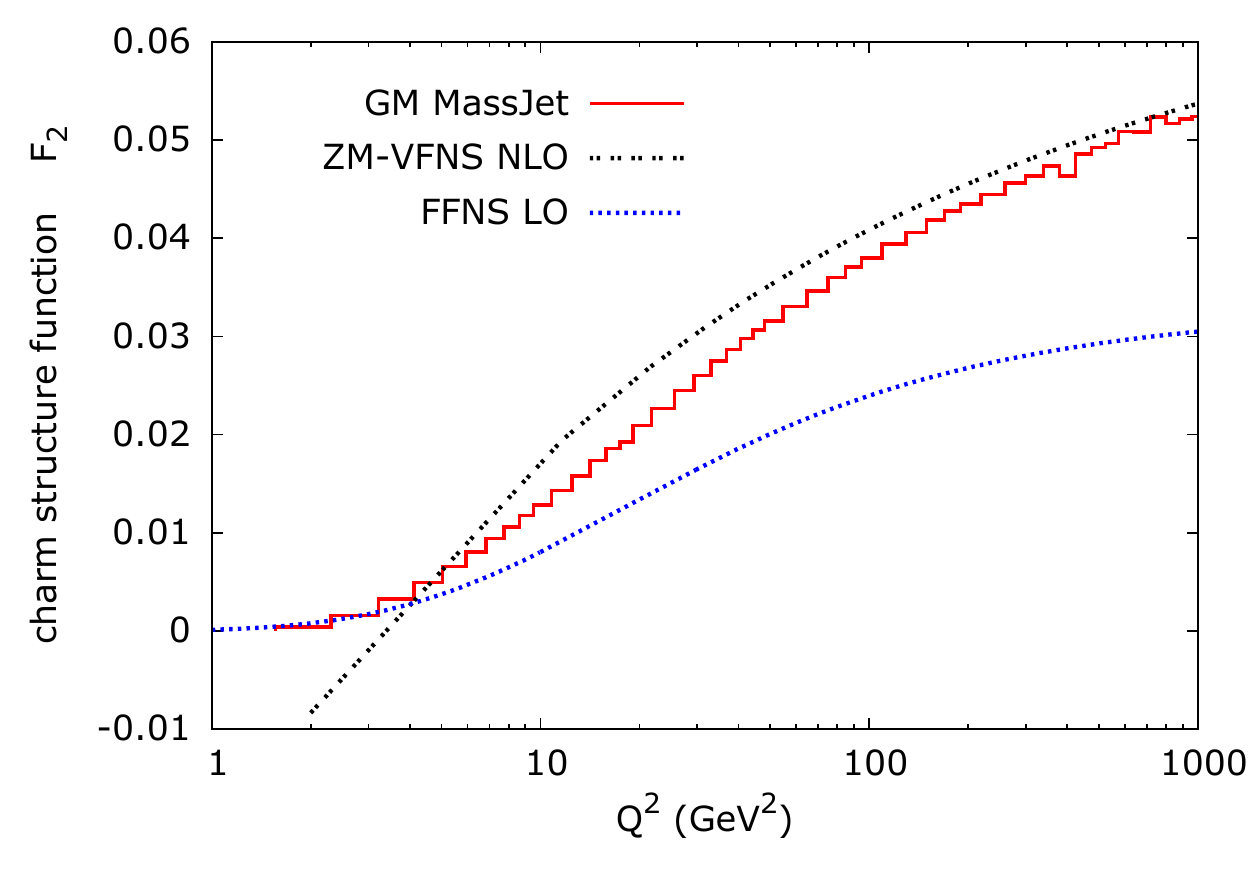}
\end{centering}
\vspace{-20pt}

\caption{\small ACOT charm structure function calculated using MC implementation of our method ($\mathtt{MassJet}$). The calculations are
done for $x_{B}=0.05$ and CTEQ5 PDFs. \label{Fig:GeneralMass_Example_2}}
\end{wrapfigure}%

The first step towards GM scheme for jets is to construct dipole functions
controlling \mbox{q-singularities} for initial state emissions. Moreover we want to have
possibly massive initial states as it is allowed by the ACOT scheme. It was partially
done in \cite{Dittmaier:1999mb} (for $\mathbf{Q}\rightarrow\mathbf{Q}g$
splitting), while in \cite{Catani:2002hc} the splitting processes with heavy quarks are considered in the final states only.

Let us look at the particular example. Consider the initial state
$g\rightarrow\mathbf{Q}\overline{\mathbf{Q}}$ splitting. Let us assign the
momentum $p_{a}$ to the gluon, $p_{i}$ to the emitted final state quark (or anti-quark), 
and $p_{j}$ to the spectator. Using these, we construct new momenta which enter $\mathcal{M}_{n,a}$ 
in (\ref{eq:dipole}): $\tilde{p}_{j}^{\mu}=\tilde{w}\left(p_{i}^{\mu}+p_{j}^{\mu}\right)-\tilde{u}p_{a}^{\mu}$
becomes a new final state and $\tilde{p}_{\underline{ai}}^{\mu}=\left(\tilde{w}-1\right)\left(p_{i}^{\mu}+p_{j}^{\mu}\right)-\left(\tilde{u}-1\right)p_{a}^{\mu}$
becomes a new initial state. The variables $\tilde{u}$, $\tilde{w}$
can be determined from on-shell conditions for $\tilde{p}_{j}$ and
$\tilde{p}_{\underline{ai}}$. Our dipole splitting function reads\begin{equation}
\hat{V}_{g\rightarrow\mathbf{Q}\overline{\mathbf{Q}},\, j}=8\pi\mu_{\mathrm{r}}^{2\varepsilon}\alpha_{s}T_{R}\left[1-\frac{1}{1-\varepsilon}\,\left(2\tilde{u}\left(1-\tilde{u}\right)-\frac{\left(1-\tilde{u}\right)m_{\mathbf{Q}}^{2}}{p_{i}\cdot p_{a}}\right)\right],\label{eq:Dipsplit_IEFS_g_QQ_V}\end{equation}
where $\mu_{r}$ is a mass scale needed in $D=4-2\varepsilon$ dimensions.
In this case $\hat{V}$ is just diagonal in helicity space.

Consider next the integral of (\ref{eq:Dipsplit_IEFS_g_QQ_V}) over
one-particle subspace. It can be convenietly expressed in terms of
rescaled masses $\eta_{l}^{2}=m_{l}^{2}/2\tilde{p}_{j}\cdot p_{a}$
for some parton $l$. In the limit of small $\eta_{\mathbf{Q}}$
we get\begin{equation}
\int d\phi\,\hat{V}_{g\rightarrow\mathbf{Q}\overline{\mathbf{Q}},\, j}\left(u\right)=\frac{\alpha_{s}}{2\pi}\Bigg[P_{gq}\left(u\right)\Bigg(\log\frac{u^{2}}{u+\eta_{j}^{2}}-\log\eta_{\mathbf{Q}}^{2}\Bigg)+2T_{R}\, u\left(1-u\right)\Bigg]+\mathcal{O}\left(\eta_{\mathbf{Q}}^{2}\right).\label{eq:V_gQQ_int}\end{equation}
We see that there is a term of the form $P_{gq}\left(u\right)\log\eta_{\mathbf{Q}}^{2}$
which becomes harmful when the scale becomes large (in massless case
there would be a pole $1/\varepsilon$). Similar terms appear also in
other dipoles for the initial state emissions.

\section{General mass scheme for jets}

In the spirit of the ACOT scheme, the initial state q-singularities
have to be factorized out. It is accomplished by
$\mathcal{C}_{n,a}$ term with partonic PDFs $\mathcal{F}_{ab}$ calculated in a special
way. Let us recall at this point, that the latter are defined as certain
ME of light-cone operators and can be calculated order by order using
special Feynman rules (see e.g. \cite{Collins:2011zzd}). The results contain UV 
singularities which have to be renormalized, leading to evolution
equations. For the present application we calculate $\mathcal{F}_{ab}$
to one loop with full mass dependence and renormalize them using the $\overline{\mathrm{MS}}$
scheme%
\footnote{More precisely we use the Collins-Wilczek-Zee renormalization scheme which
assumes the $\overline{\mathrm{MS}}$ above certain threshold and zero
momentum subtraction below it.%
}. Since counterterms are mass independent in this scheme, we assure that hadronic PDFs evolve according to standard massless DGLAP equations.
For instance we get \begin{gather}
\mathcal{F}_{g\mathbf{Q}}\left(z\right)=\frac{\alpha_{s}}{2\pi}\, T_{R}\,\left(1-2z\left(1-z\right)\right)\,\log\frac{\mu_{r}^{2}}{m_{\mathbf{Q}}^{2}},\\
\mathcal{\mathcal{F}}_{\mathbf{Q}g}\left(z\right)=\frac{\alpha_{s}}{2\pi}\, C_{F}\,\frac{1+\left(1-z\right)^{2}}{z}\left[\log\frac{\mu_{r}^{2}}{m_{\mathbf{Q}}^{2}}-2\log z-1\right],\\
\mathcal{\mathcal{F}}_{\mathbf{Q}\mathbf{Q}}\left(z\right)=\frac{\alpha_{s}}{2\pi}\, C_{F}\left\{ \frac{1+z^{2}}{1-z}\left[\log\frac{\mu_{r}^{2}}{m_{\mathbf{Q}}^{2}}-2\log\left(1-z\right)-1\right]\right\} _{+}.\end{gather}

We have checked that the above procedure leads to IR safe
dipoles in the limit of vanishing $\eta_{\mathbf{Q}}$. Moreover, the results
coincide with those Ref. \cite{Catani:2002hc} in the $\overline{\mathrm{MS}}$
scheme.


In order to perform numerical tests we have partially implemented
our method in a dedicated C++ program based on FOAM \cite{Jadach:2002kn}.
Using the program, we have calculated the charm structure function $F_{2}$
and compared it with semi-analytical calculation in the ACOT scheme. This exercise uses three dipoles
and two collinear subtraction terms. The virtual corrections are taken
from \cite{Kretzer:1998ju}. We find that the soft poles are indeed
cancelled by the corresponding poles coming from the integrated dipoles.
Moreover, we find agreement with the semi-analytical calculation and observe
that our result properly interpolates between the two limiting solutions of the
ZM-VFNS and FFNS schemes, as depicted in Fig. \ref{Fig:GeneralMass_Example_2}.

Let us stress that the result is obtained by a numerical integration of a fully differential cross section,
which provides a severe test on the implementation of our massive dipole formalism.


{\raggedright
\begin{footnotesize}


\bibliographystyle{DISproc}
\bibliography{Kotko_Piotr}
\end{footnotesize}
}


\end{document}